\documentclass{moriond}

\def\be{\begin{equation}}
\def\ee{\end{equation}}
\def\bea{\begin{eqnarray}}
\def\eea{\end{eqnarray}}

\usepackage{amsmath,amssymb,mathtools}
\usepackage{graphicx}
\usepackage{booktabs,cellspace}
\usepackage{color}
\usepackage{hyperref}
\usepackage{xspace}
\newcommand{\as}{\ensuremath{\alpha_s}\xspace}

\newcommand{\pb}{\ensuremath{\mathrm{pb}}\xspace}

\newcommand{\GeV}{\ensuremath{\mathrm{GeV}}\xspace}
\newcommand{\TeV}{\ensuremath{\mathrm{TeV}}\xspace}
\newcommand{\mll}{\ensuremath{{M_{{\ell\ell}}}}\xspace}
\newcommand{\pt}{\ensuremath{{p_T^{{\ell\ell}}}}\xspace}

\newcommand{\ntlo}{\text{N${}^3$LO}\xspace}
\newcommand{\ntll}{\text{N${}^3$LL}\xspace}

\newcommand{\RadISH}{\texttt{RadISH}\xspace}
\newcommand{\NNLOJET}{\texttt{NNLOJET}\xspace}

\newcommand\blfootnote[1]{%
  \begingroup
  \renewcommand\thefootnote{}\footnote{#1}%
  \addtocounter{footnote}{-1}%
  \endgroup
}

\newcommand{\Photo}{}

\begin{document}
\vspace*{4cm}
\title{Theory uncertainties in the fiducial Drell--Yan cross section and distributions}

\author{
Xuan Chen$^{1,2}$,
  Thomas Gehrmann$^3$,
  Nigel Glover$^4$,
  Alexander Huss$^5$,
  Pier Francesco Monni$^5$,\\
  Emanuele Re$^{6,7}$,
  Luca Rottoli$^3$$^\dagger$\blfootnote{$^\dagger$ Speaker.},
  Paolo Torrielli$^8$
  }

\address{
$^1$ Institute for Theoretical Physics, Karlsruhe Institute of Technology, 76131 Karlsruhe, Germany\\
$^2$ Institute for Astroparticle Physics, Karlsruhe Institute of Technology, 76344 Eggenstein-Leopoldshafen, Germany\\
$^3$ Department of Physics, University of Z\"urich, CH-8057 Z\"urich, Switzerland\\
$^4$ Institute for Particle Physics Phenomenology, Physics Department, Durham University, Durham DH1 3LE, UK\\
$^5$ CERN, Theoretical Physics Department, CH-1211 Geneva 23, Switzerland\\
$^6$ Dipartimento di Fisica G. Occhialini,\\
  U2, Universit\`a degli Studi di Milano-Bicocca and INFN,
  Sezione di Milano-Bicocca,\\
  Piazza della Scienza, 3, 20126 Milano, Italy\\
 $^7$ LAPTh, Universit\'e Grenoble Alpes, Universit\'e Savoie Mont Blanc, CNRS, F-74940 Annecy, France\\
$^8$ Dipartimento di Fisica and Arnold-Regge Center,
  Universit\`a di Torino and INFN, Sezione di Torino,
  Via P. Giuria 1, I-10125, Turin, Italy
}

\maketitle

\abstracts{ In these proceedings we study various sources of
  theoretical uncertainty in the Drell--Yan $\pt$ spectrum focussing
  on the $\pt \lesssim 100\, \GeV$ region.  We consider several
  perturbative aspects related to the choice of the scale setting
  adopted in resummed calculations, and we assess their impact on the
  theoretical prediction both for the differential $\pt$ spectrum and
  for the \ntlo\ fiducial cross section.
  For both quantities, we find the results obtained with the different
  setups to be compatible with each other within the quoted
  uncertainty, highlighting the robustness of the theoretical
  prediction.
  In all cases, the experimental LHC data for the $\pt$ spectrum is
  well described by our calculation.}

\paragraph{Introduction.---}%
The theoretical description of the Drell--Yan $\pt$ spectrum is among
the most challenging tasks in collider physics at present, due to the
outstanding accuracy reached by the experimental measurements.
In a recent article~\cite{Chen:2022cgv}, we have presented the state
of the art, N$^3$LO calculation of the fiducial Drell--Yan cross
section and its leptonic distributions such as $\pt$, also studying
the effect of the inclusion of QCD resummation of large logarithms of
$\mll/\pt$, with $\mll$ being the Drell--Yan pair invariant mass.
Our findings indicate that a first-principle calculation using
perturbative QCD methods describes well the experimental data for the
differential $\pt$ distribution measured in the regime $\mll\sim M_Z$,
with the exception of the very small $\pt$ region where a
phenomenological modelling of non-perturbative effects is needed.

The high accuracy of the theoretical calculation requires a careful
estimate of the associated uncertainties, which are below $\pm 5\%$
for $\pt \lesssim 100\, \GeV$.
In these proceedings we briefly review the calculation of Ref.~\cite{Chen:2022cgv}
and we discuss various sources of theoretical uncertainty that are relevant for the
description of the Drell--Yan $\pt$ spectrum.
We focus our discussion on aspects particularly relevant in the
$\pt \lesssim 100\, \GeV$ region: the resummation of logarithmic corrections and the matching of resummation with the fixed order prediction.

The prediction of Ref.~\cite{Chen:2022cgv} for the differential $\pt$ spectrum is based
on a combination of a resummed calculation at N$^3$LL (including
constant terms up to ${\cal O}(\alpha_s^3)$) obtained with
\RadISH~\cite{Monni:2016ktx,Bizon:2017rah,Monni:2019yyr} with the NNLO
calculation obtained with
\NNLOJET~\cite{Gehrmann-DeRidder:2015wbt,Gehrmann-DeRidder:2016cdi,Gehrmann-DeRidder:2016jns}.
The formulation adopted in the \RadISH code is based on a
momentum-space formalism and does
not introduce any modelling of non-perturbative effects. Instead, the Landau singularity is regularised by freezing the running of the strong coupling constant at scales of the
order of $0.5\,\GeV$ and that of the parton distribution functions
(PDFs) at the extraction scale of the set adopted, which in this case
corresponds to $1.65\,\GeV$~\cite{Ball:2021leu}.
We will comment further on the prescription used to freeze the PDFs below. 
This prescription leads to effects in the calculation in the first two
bins of the $\pt$ distribution where non perturbative (NP) dynamics
becomes relevant.
In this region a realistic modelling of NP corrections is therefore necessary.

In the following, we study the impact on the theoretical uncertainties
of various sources of higher-order corrections,
related to the central scale setting used in our
predictions.

\paragraph{Computational setup.---}%
Throughout this note, we will consider proton--proton collisions at a
centre-of-mass energy $\sqrt{s}=13\,\TeV$, and we adopt the
\texttt{NNPDF4.0} parton densities~\cite{Ball:2021leu} at NNLO with
$\as(M_Z)=0.118$, whose scale evolution is performed with
\texttt{LHAPDF}~\cite{Buckley:2014ana} and
\texttt{Hoppet}~\cite{Salam:2008qg}, correctly accounting for
heavy-quark thresholds.
We set the central factorisation and renormalisation scales to
$\mu_F = \mu_R = \sqrt{{M_{\ell \ell}}^2 + {p_T^{\ell  \ell}}^2}$.
We adopt the $G_\mu$ scheme with the following EW parameters taken
from the PDG~\cite{ParticleDataGroup:2018ovx}: $M_Z = 91.1876\,\GeV$,
$M_W = 80.379\,\GeV$, $\Gamma_Z = 2.4952\,\GeV$,
$\Gamma_W = 2.085\,\GeV$, and
$G_F=1.1663787\times 10^{-5}\,\GeV^{-2}$.
We consider a fiducial volume~\cite{ATLAS:2019zci} in which the
leptonic invariant mass window is constrained to be
$66\,\GeV < \, \mll \,< 116\,\GeV$ and the lepton rapidities are
confined to $|\eta^{\ell^\pm}| < 2.5$. The transverse momenta of the
two leptons are required to
satisfy $|\vec{p}_{T}^{~\ell^\pm}| > 27\,\GeV$.

\paragraph{Resummation scheme and scale setting for the ${\cal O}(\alpha_s^3)$ constant
  terms.---}%
We start by discussing the impact of the choice of the strong coupling scale in the ${\cal O}(\alpha_s^3)$ constant terms in the resummation formula.
Here we deliberately use a very schematic and simplistic language to
introduce how they arise in the resummed calculation. An appropriate
discussion of these terms and their structure within the \RadISH
framework is reported in Ref.~\cite{Re:2021con}, where the scale
setting adopted in Ref.~\cite{Chen:2022cgv} is discussed in detail.
Schematically, one can parametrise the perturbative logarithmic
counting for the resummed cumulative cross section considered here as
\begin{align}
\label{eq:master}
 {\cal C}(\alpha_s(\mll),\alpha_s(\mu))\, \exp\{\sum_{i=-1}^2
            \alpha_s^{i}\,h_{i+2}(\alpha_s L) + \dots\},
\end{align}
where $\alpha_s \equiv \alpha_s(\mll)$ and $L$ denotes the large
logarithms which are resummed in the $\pt\ll \mll$ regime. The
function ${\cal C}(\alpha_s(\mll),\alpha_s(\mu))$ encodes constant
contributions that survive in the $\pt\to 0$ limit. It admits a
perturbative expansion in powers of the strong coupling $\alpha_s$,
which is needed up to three loops
(${\cal
  O}(\alpha_s^3)$)~\cite{Gehrmann:2010ue,Li:2016ctv,Luo:2019szz,Ebert:2020yqt}
in order to achieve, together with the functions $h_i(\alpha_s L)$,
the accuracy of the predictions of Ref.~\cite{Chen:2022cgv} in the
regime $\alpha_s L\sim 1$ and $\alpha_s \ll 1$.

We first turn our focus on the scale of the coupling
constant in the perturbative expansion of
${\cal C}(\alpha_s(\mll),\alpha_s(\mu))$. At each order
${\cal O}(\alpha_s^i)$ this receives contributions both from terms
evaluated at $\alpha_s^i(\mll)$ and from terms evaluated at
$\alpha_s^i(\mu)$, where the scale $\mu\ll \mll$ is of the order of
the transverse momentum of the QCD radiation probed in the $\pt \to 0$
limit. The precise value of this scale depends on the resummation
formalism. Within \RadISH this is set as $\mu\sim k_{t1}$, with
$ k_{t1}$ being the transverse momentum of the hardest initial-state
radiation, while in an impact-parameter approach this is set as
$\mu\sim 1/b$, with $b$ being the impact parameter (see
e.g. Refs.~\cite{Collins:1984kg,Catani:2000vq,Becher:2010tm}). These
two scales are of the same order and are related by a Bessel integral
transform~\cite{Bizon:2017rah}.

In a direct QCD formulation of $\pt$ resummation, the separation of
terms evaluated at $\alpha_s^i(\mll)$ and those evaluated at
$\alpha_s^i(\mu)$ specifies the so-called \textit{resummation
  scheme}~\cite{Catani:2000vq}, and only the combination of both
factors in Eq.~\eqref{eq:master} is resummation-scheme
invariant. More precisely, if we denote by ${\cal C}^{(i)}$ the
${\cal O}(\alpha_s^{i})$ term of the perturbative expansion of
${\cal C}(\alpha_s(\mll),\alpha_s(\mu))$, only the combination of
${\cal C}^{(i)}$ and $h_{i+2}(\alpha_s L)$ is resummation-scheme
invariant.
Consequently, a change in the renormalisation scale in the
${\cal O}(\alpha_s^3)$ constant terms ${\cal C}^{(3)}$ will affect the
form of the correction $\alpha_s^3 h_{5}(\alpha_s L)$ in
Eq.~\eqref{eq:master}, which is a genuine N$^4$LL correction and hence
beyond the perturbative accuracy of the calculation discussed here.

The prediction presented in Ref.~\cite{Chen:2022cgv} evaluates the
${\cal O}(\alpha_s^3)$ terms with $\alpha_s(\mll)$; it is however possible
to evaluate the terms of hard-virtual origin 
in ${\cal C}^{(3)}$ at
$\alpha_s(\mll)$, while those of soft and/or collinear origin are
evaluated at $\mu\sim k_{t1}\ll \mll$.
The difference between the two prescriptions is, as explained above,
subleading in the perturbative order of the calculation. As such, one
expects it to be compatible within the quoted perturbative
uncertainties.

\begin{figure}[t!]
  \centering
  \includegraphics[width=0.49\linewidth]{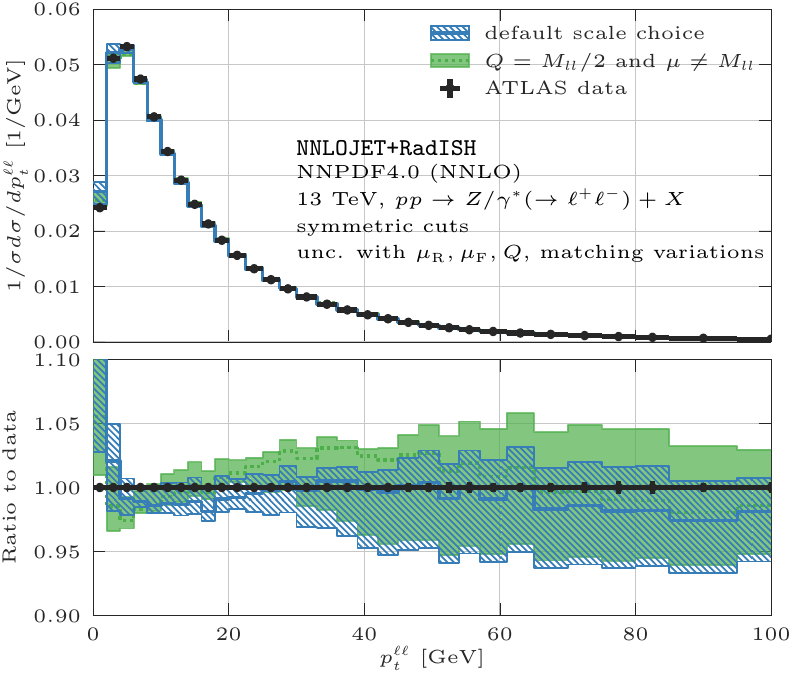}
 \includegraphics[width=0.49\linewidth]{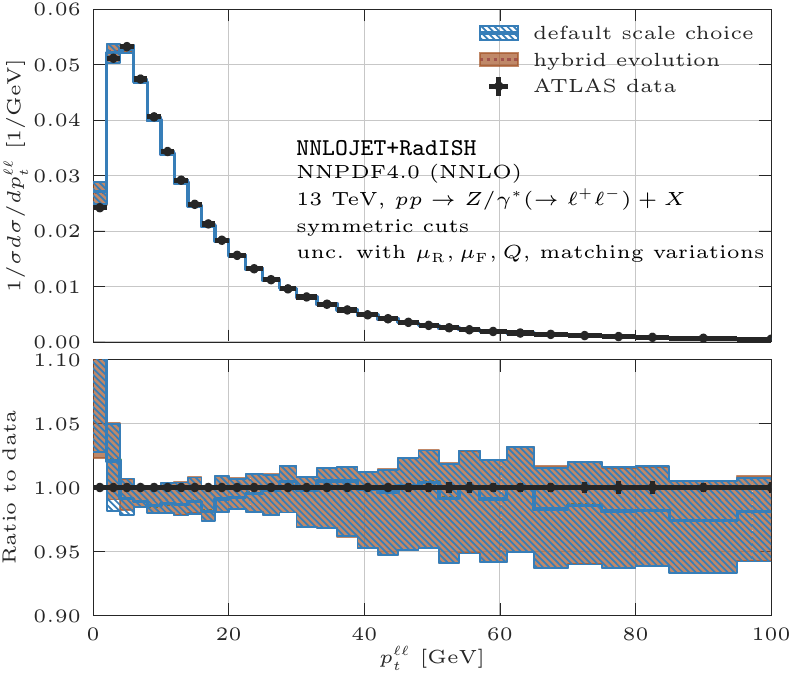}
  \caption{\label{fig:running} Left panel: fiducial \pt distribution at
    \ntlo{}+\ntll in the default scale setup of
    Ref.~\protect\cite{Chen:2022cgv} (blue, hatched) and evaluating the
    ${\cal O}(\alpha_s^3)$ constant terms of soft and collinear origin
    at $\mu=k_{t1}$ (green, solid). See Ref.~\protect\cite{Re:2021con} for a
    more detailed discussion. Right panel: Fiducial \pt distribution at
    \ntlo{}+\ntll in the default scale setup of
    Ref.~\protect\cite{Chen:2022cgv} (blue, hatched) and using the hybrid
    evolution for the parton densities below the extraction scale
    (brown, solid).}
\end{figure}

A study of the difference between the two scale settings in the
\RadISH{}+\NNLOJET prediction was presented in Ref.~\cite{Re:2021con}.
The difference between the two scale settings is shown in
Fig.~\ref{fig:running} (left), where the blue, hatched band represents our
default setup used in Ref.~\cite{Chen:2022cgv}, and the green, solid
band shows the result with the constant terms of soft and/or collinear
origin evaluated at the scale $\mu=k_{t1}$ (labelled with
$\mu\ne \mll$ in the plot) up to ${\cal O}(\alpha_s^3)$.
The uncertainties are still estimated as outlined in
Ref.~\cite{Chen:2022cgv}. In particular, this prescription includes a
very conservative estimate of the matching uncertainty, which is
obtained by taking the envelope of the uncertainty bands obtained with
\textit{four} different matching schemes, for a total of 36 variations
(9 scale variations per scheme). This conservative approach is taken given the
level of precision that is reached by the
perturbative calculation.
Fig.~\ref{fig:running} shows that the change in the scale $\mu$ leads
to a distortion of the spectrum in such a way that it becomes softer
at small $\pt$ and slightly harder for $\pt > 10\,\GeV$. The resulting
distribution in Fig.~\ref{fig:running} is still compatible with the
data and with our default setup within uncertainties, in line with the
fact that it corresponds to a subleading logarithmic correction.
An exception is the region between $[20,40]\,\GeV$ where the central
value of the green band lies outside the error band (blue) of our
default setup, suggesting that one may adopt a slightly more
conservative error estimate that includes the central curve of the
green band in the envelope that defines the theory uncertainty.
%


\paragraph{Evolution of parton densities and freezing.---}%
We now briefly discuss how the freezing of the parton densities at the
extraction scale of $1.65\,\GeV$~\cite{Ball:2021leu} impacts our
prediction.
As an alternative prescription, instead of freezing the parton
densities at $Q_0=1.65\,\GeV$, we evolve backward from this scale down
to $0.5\,\GeV$ taking correctly into account the charm-quark mass
threshold. This ensures that possible small artefacts related to the
freezing of the parton densities are pushed to the very small $\pt$
region.
%
Fig.~\ref{fig:running} (right) shows the comparison of the default setup of
Ref.~\cite{Chen:2022cgv} to the prediction obtained with the above
treatment of the parton distribution functions, that we label as
\textit{hybrid} in the plot.
%
%
As it can be appreciated from the figure, the freezing only modifies
the prediction for this setup at very small $\pt$ values, in a way
that is fully compatible with our estimate of the theory
uncertainties.

\paragraph{The resummation scale.---}%
Finally, we discuss another relevant aspect in the scale setting used
in the \RadISH calculations, which concerns the
value of the hard scale of the process, of the order of $\mll$.
This hard scale is set to $\mll/2$ in the \RadISH predictions of
Ref.~\cite{Chen:2022cgv}.
An associated variation of this perturbative scale (as well as of the
other perturbative scales) is encoded in the estimate of the
perturbative uncertainties.
In the \RadISH formalism, this scale enters in the form of a
resummation scale $Q$. This can be introduced by decomposing the
resummed logarithm $L$ as~\cite{Bizon:2018foh,Re:2021con}
\begin{equation}
L = \tilde{L} + \ln\frac{\mll}{Q}\,,
\end{equation}
and re-expanding $L$ about $\tilde{L}$ while neglecting subleading
(N$^4$LL) corrections. This is motivated by the fact that in the
$\pt\to 0$ limit one has $\tilde{L} \gg \ln\frac{\mll}{Q}$.
A variation of the scale $Q$ (commonly by a factor of two about its
central value) then probes the size of subleading logarithmic
corrections in the uncertainty estimate.
At the same time, the logarithm $\tilde{L}$ is switched off for
$\pt \gtrsim Q$ with a smooth
deformation~\cite{Bizon:2018foh,Re:2021con} that introduces power
corrections of order $(\pt/Q)^{p-1}$ (we choose the parameter $p=6$)
in the $\pt$ differential distributions. These facilitate the matching
of the resummed result to the fixed order calculation and appear only
at subleading orders with respect to the nominal result, that is at
${\cal O}(\alpha_s^4)$. In this way, the scale $Q$ also takes the role
of the scale at which resummation effects are switched off in the
$\pt$ distribution.
Choosing $Q=\mll$ as a central scale implies that a variation of $Q$
in the assessment of the theory uncertainty will induce residual
resummation effects up to $\pt\sim 2\mll$, which is a rather high
scale. For this reason, our default setup is to vary $Q$ about its
central value $\mll/2$, above which QCD is well described by
fixed-order perturbation theory.

\begin{figure}[t!]
  \centering
  \includegraphics[width=0.49\linewidth]{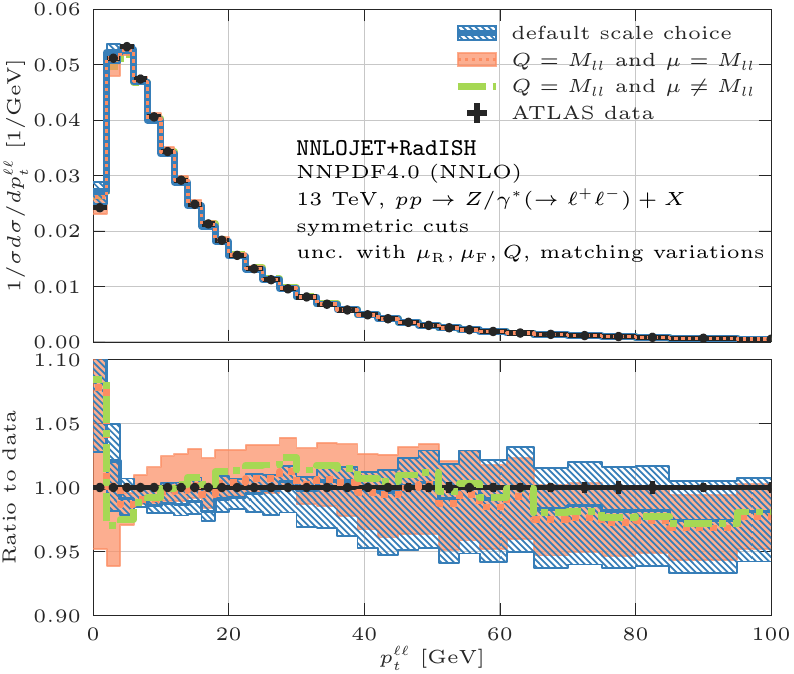}
  \caption{\label{fig:resscale} Fiducial \pt distribution at
    \ntlo{}+\ntll in the default scale setup of
    Ref.~\protect\cite{Chen:2022cgv} (blue, hatched) and setting the central
    value of the resummation scale at $Q=\mll$ (red, solid). The
    green-dashed curve indicates the central prediction with $Q=\mll$
    and $\mu\ne \mll$.}
\end{figure}

Nevertheless, in the following we study the difference between the
two scale settings. For this reason, in
Fig.~\ref{fig:resscale} we show the comparison between the
\RadISH{}+\NNLOJET prediction using either $Q=\mll/2$ (our default,
given by the blue, hatched band) or $Q=\mll$ (given by the red, solid
band) as a central scale.
In the latter, we also adopt the hybrid treatment of the evolution of
parton distribution functions (PDFs) discussed above, to avoid any
interplay between the PDFs freezing scale and the perturbative
prediction in the small $\pt$ region.
The uncertainties in the red band of Fig.~\ref{fig:resscale} are
estimated as done in Ref.~\cite{Re:2021con}, by performing scale
variations within three matching schemes~\footnote{In this case, due
  to the high resummation scale, we only consider the three matching
  schemes involving a matching factor (see
  Ref.~\cite{Re:2021con}). The corresponding uncertainty thus consists
  of 27 variations.} that ensure that the resummation is switched off
at $\pt\sim \mll$.
We observe that the prediction with $Q=\mll$ as a central scale
receives a shift in the upward (downward) direction for
$\pt> (<) \,10\,\GeV$. Moreover, the perturbative uncertainty grows
with this different resummation scale, and the prediction remains
compatible with the experimental data.

Finally, it is instructive to adopt the three modifications to our
default setting discussed in this note together. That is, we use
$\mu\ne \mll$ as outlined in the previous section together with
$Q=\mll$ in our prediction. The result is displayed by the dot-dashed
light-green line in Fig.~\ref{fig:resscale}, and it is entirely within
the (red) uncertainty band of the $Q=\mll$ result as one would expect
from a subleading effect.

\paragraph{Impact on the fiducial cross section.---}%
As a last step, we also discuss the impact of the different scale
setting discussed in this note on the fiducial cross section at
N$^3$LO+N$^3$LL presented in Ref.~\cite{Chen:2022cgv}.
This reference quotes $726.2(1.1)^{+1.07\%}_{-0.77\%}\,\pb$ for the
fiducial cross section within symmetric cuts, compatible with the
N$^3$LO result of $722.9(1.1)^{+0.68\%}_{-1.09\%} \pm 0.9\,\pb$,
computed in the same article.
The central value of the N$^3$LO+N$^3$LL cross section obtained within
the setup shown in the green band of Fig.~\ref{fig:running} (left) (namely
with $\mu\ne\mll$) is $725.0(1.1)\,\pb$, while the central value
corresponding to the red band of Fig.~\ref{fig:resscale} is
$723.8(1.1)\,\pb$. Both values are well within the scale uncertainty
of the N$^3$LO+N$^3$LL calculation performed in
Ref.~\cite{Chen:2022cgv}, which confirms the robustness of this
prediction for the fiducial cross section.

\paragraph{Conclusions.---}%

\sloppy
In these proceedings we have discussed the dependence of the
\RadISH{}+\NNLOJET{} predictions of Ref.~\cite{Chen:2022cgv} upon
different choices for the perturbative scale setting.
We observed that all setups yield results which are compatible with
each other within the quoted uncertainties. This indicates that the
latter uncertainty is reliable, although a more conservative estimate
could be envisaged by taking into account the spectrum of variations
considered here.
All of our predictions agree well with the experimental LHC data,
possibly with the exclusion of the very small $\pt$ region, which
requires a careful assessment of non-perturbative effects.

We have also examined the impact of the different setups on the
fiducial cross section at \ntlo{}+\ntll{}, finding in all cases that
the effect of the different scale choices is well within the
perturbative uncertainty band obtained in Ref.~\cite{Chen:2022cgv},
highlighting its robustness.

\paragraph{Acknowledgments.---}

We thank the organisers of the 56$^{\rm{th}}$ Rencontres de Moriond 2022.
This work has received funding from the Deutsche Forschungsgemeinschaft (DFG, German Research Foundation) under grant 396021762-TRR 257, from the Swiss National Science Foundation (SNF) under contracts
PZ00P2$\_$201878, 200020$\_$188464 and 200020$\_$204200,
from the UK Science and Technology Facilities Council (STFC) through grant ST/T001011/1, from the Italian Ministry of University and Research (MIUR) through grant PRIN 20172LNEEZ,
and from the European Research Council (ERC) under the European Union's Horizon 2020 research and innovation programme grant agreement 101019620 (ERC Advanced Grant TOPUP).

\bibliography{letter}

\begin{thebibliography}{10}

\bibitem{Chen:2022cgv}
X.~Chen, T.~Gehrmann, E.~W.~N. Glover, A.~Huss, P.~Monni, E.~Re, L.~Rottoli,
  and P.~Torrielli.
\newblock {Third-Order Fiducial Predictions for Drell-Yan Production at the
  LHC}.
\newblock {\em Phys. Rev. Lett.}, 128(25):252001, 2022.

\bibitem{Monni:2016ktx}
Pier~Francesco Monni, Emanuele Re, and Paolo Torrielli.
\newblock {Higgs Transverse-Momentum Resummation in Direct Space}.
\newblock {\em Phys. Rev. Lett.}, 116(24):242001, 2016.

\bibitem{Bizon:2017rah}
Wojciech Bizon, Pier~Francesco Monni, Emanuele Re, Luca Rottoli, and Paolo
  Torrielli.
\newblock {Momentum-space resummation for transverse observables and the Higgs
  p$_{\perp}$ at N$^{3}$LL+NNLO}.
\newblock {\em JHEP{}}, 02:108, 2018.

\bibitem{Monni:2019yyr}
Pier~Francesco Monni, Luca Rottoli, and Paolo Torrielli.
\newblock {Higgs transverse momentum with a jet veto: a double-differential
  resummation}.
\newblock {\em Phys. Rev. Lett.}, 124(25):252001, 2020.

\bibitem{Gehrmann-DeRidder:2015wbt}
A.~Gehrmann-De~Ridder, T.~Gehrmann, E.~W.~N. Glover, A.~Huss, and T.~A. Morgan.
\newblock {Precise QCD predictions for the production of a Z boson in
  association with a hadronic jet}.
\newblock {\em Phys. Rev. Lett.}, 117(2):022001, 2016.

\bibitem{Gehrmann-DeRidder:2016cdi}
Aude Gehrmann-De~Ridder, T.~Gehrmann, E.~W.~N. Glover, A.~Huss, and T.~A.
  Morgan.
\newblock {The NNLO QCD corrections to Z boson production at large transverse
  momentum}.
\newblock {\em JHEP{}}, 07:133, 2016.

\bibitem{Gehrmann-DeRidder:2016jns}
A.~Gehrmann-De~Ridder, T.~Gehrmann, E.~W.~N. Glover, A.~Huss, and T.~A. Morgan.
\newblock {NNLO QCD corrections for Drell-Yan $p_T^Z$ and $\phi^*$ observables
  at the LHC}.
\newblock {\em JHEP{}}, 11:094, 2016.
\newblock [Erratum: JHEP 10, 126 (2018)].

\bibitem{Ball:2021leu}
Richard~D. Ball et~al.
\newblock {The Path to Proton Structure at One-Percent Accuracy}.
\newblock 9 2021.

\bibitem{Buckley:2014ana}
Andy Buckley, James Ferrando, Stephen Lloyd, Karl Nordstr\"om, Ben Page, Martin
  R\"ufenacht, Marek Sch\"onherr, and Graeme Watt.
\newblock {LHAPDF6: parton density access in the LHC precision era}.
\newblock {\em Eur. Phys. J. C}, 75:132, 2015.

\bibitem{Salam:2008qg}
Gavin~P. Salam and Juan Rojo.
\newblock {A Higher Order Perturbative Parton Evolution Toolkit (HOPPET)}.
\newblock {\em Comput. Phys. Commun.}, 180:120--156, 2009.

\bibitem{ParticleDataGroup:2018ovx}
M.~Tanabashi et~al.
\newblock {Review of Particle Physics}.
\newblock {\em Phys. Rev. D}, 98(3):030001, 2018.

\bibitem{ATLAS:2019zci}
Georges Aad et~al.
\newblock {Measurement of the transverse momentum distribution of
  Drell\textendash{}Yan lepton pairs in proton\textendash{}proton collisions at
  $\sqrt{s}=13$ TeV with the ATLAS detector}.
\newblock {\em Eur. Phys. J. C}, 80(7):616, 2020.

\bibitem{Re:2021con}
Emanuele Re, Luca Rottoli, and Paolo Torrielli.
\newblock {Fiducial Higgs and Drell-Yan distributions at N$^3$LL$^\prime$+NNLO
  with RadISH}.
\newblock {\em JHEP{}}, 09:108, 2021.

\bibitem{Gehrmann:2010ue}
T.~Gehrmann, E.~W.~N. Glover, T.~Huber, N.~Ikizlerli, and C.~Studerus.
\newblock {Calculation of the quark and gluon form factors to three loops in
  QCD}.
\newblock {\em JHEP{}}, 06:094, 2010.

\bibitem{Li:2016ctv}
Ye~Li and Hua~Xing Zhu.
\newblock {Bootstrapping Rapidity Anomalous Dimensions for Transverse-Momentum
  Resummation}.
\newblock {\em Phys. Rev. Lett.}, 118(2):022004, 2017.

\bibitem{Luo:2019szz}
Ming-xing Luo, Tong-Zhi Yang, Hua~Xing Zhu, and Yu~Jiao Zhu.
\newblock {Quark Transverse Parton Distribution at the
  Next-to-Next-to-Next-to-Leading Order}.
\newblock {\em Phys. Rev. Lett.}, 124(9):092001, 2020.

\bibitem{Ebert:2020yqt}
Markus~A. Ebert, Bernhard Mistlberger, and Gherardo Vita.
\newblock {Transverse momentum dependent PDFs at N$^3$LO}.
\newblock {\em JHEP{}}, 09:146, 2020.

\bibitem{Collins:1984kg}
John~C. Collins, Davison~E. Soper, and George~F. Sterman.
\newblock {Transverse Momentum Distribution in Drell-Yan Pair and W and Z Boson
  Production}.
\newblock {\em Nucl. Phys. B}, 250:199--224, 1985.

\bibitem{Catani:2000vq}
Stefano Catani, Daniel de~Florian, and Massimiliano Grazzini.
\newblock {Universality of nonleading logarithmic contributions in transverse
  momentum distributions}.
\newblock {\em Nucl. Phys. B}, 596:299--312, 2001.

\bibitem{Becher:2010tm}
Thomas Becher and Matthias Neubert.
\newblock {Drell-Yan Production at Small $q_T$, Transverse Parton Distributions
  and the Collinear Anomaly}.
\newblock {\em Eur. Phys. J. C}, 71:1665, 2011.

\bibitem{Bizon:2018foh}
Wojciech Bizo\'n, Xuan Chen, Aude Gehrmann-De~Ridder, Thomas Gehrmann, Nigel
  Glover, Alexander Huss, Pier~Francesco Monni, Emanuele Re, Luca Rottoli, and
  Paolo Torrielli.
\newblock {Fiducial distributions in Higgs and Drell-Yan production at
  N$^{3}$LL+NNLO}.
\newblock {\em JHEP{}}, 12:132, 2018.

\end{thebibliography}

\end{document}